\documentclass[apjl, iop]{emulateapj}
\usepackage{amsmath}
\usepackage{amssymb}
\usepackage{amstext}
\usepackage{apjfonts}
\usepackage{graphicx}
\usepackage{epsfig}

\begin{document}

\shorttitle{Single-degenerate SN Ia without hydrogen}
\shortauthors{Stephen Justham}
\submitted{Submitted to ApJL on 2 August 2010. This revision accepted 25 February 2011.}
\title{Single-degenerate type Ia supernovae without hydrogen contamination}

\author{Stephen Justham}   

\affil{The Kavli Institute for Astronomy and Astrophysics, Peking University, Beijing, China}
\email{sjustham@pku.edu.cn}

\begin{abstract}
The lack of hydrogen in spectra of type Ia supernovae (SN Ia) is often seen as troublesome for single-degenerate (SD) progenitor models. We argue that, since continued accretion of angular momentum can prevent explosion of the white dwarf, it may be natural for the donor stars in SD progenitors of SN Ia to exhaust their envelopes and shrink rapidly before the explosion. This outcome seems most likely for SD SN Ia progenitors where mass-transfer begins from a giant donor star, and might extend to other SD systems. Not only is the amount of hydrogen left in such a system below the current detection limit, but the donor star is typically orders of magnitude smaller than its Roche lobe by the point when a SD SN Ia occurs, in which case attempts to observe collisions between SN shocks and giant donor stars seem unlikely to succeed.  We consider the constraints on this model from the circumstellar structures seen in spectra of SN 2006X and suggest a novel explanation for the origin of this material. 
\end{abstract}


\section{Introduction}
\label{sec:intro}


The long debate over the progenitors of type Ia supernovae (SN Ia) has gained in urgency partly due to the use of SN Ia observations in cosmology, since it is somewhat embarrassing to use SN Ia for precision measurements when we don't properly understand how they are produced. The two broad classes of progenitor models are the `single-degenerate' (SD) and `double-degenerate' (DD) scenarios. In the former case, a carbon-oxygen white dwarf (CO WD) gains mass from a non-degenerate companion until the WD reaches the point of explosive carbon burning \citep{WhelanIben1973}. In the latter, the merger of two WDs leads to the SN Ia \citep{Webbink1984, IbenTutukov1984}.\footnote{Clearly, diverse progenitors may help explain the diversity of SN Ia.}

Within each of those cateogories lies a variety of possibilities. The favoured SD systems include those with donors on the main sequence (MS) or the subgiant branch \citep[collectively known as the supersoft channel; ][]{vdH+1992, Rappaport+1994, Li+vdH1997, Langer+2000, HanPhP2004, IvanovaTaam2004} and those with red-giant (RG) donors \citep{Hachisu+1996, Hachisu+1999, HachisuKato2001}. A third option, in which the donor stars are helium-rich \citep{IbenTutukov1991} has the advantage of avoiding hydrogen contamination; however, it cannot produce a large fraction of SN Ia \citep{WangHan2010}. 

The known WDs closest to the Chandrasekhar limit are in SD systems.  The WD in each of the novae U Sco, RS Oph, T CrB, and V445 Pup is inferred to have a mass very close to the Chandrasekhar limit.\footnote{Though we cannot tell whether the WD in any of these systems is a CO WD or an ONeMg WD.}
Furthermore, the observations of \cite{Patat+Sci07} suggest that the SN Ia 2006X came from a SD progenitor (see also section \ref{sec:06X}). There are also indications that the former donor stars in SD systems have been observed: in populations of unusual WDs \citep{Hansen03, Justham+09} and in individual systems (\citealt{Ruiz-Lapuente+2004}; but see, e.g., \citealt{Kerzendorf+2009}).


The strongest arguments against the SD channel are the apparent lack of sufficient progenitor systems and the observational lack of hydrogen in SN Ia spectra.  Binary population synthesis generally struggles to produce sufficient SD SN Ia to explain the empirical SN Ia rate \citep[see, e.g.,][though there are considerable theoretical uncertainties]{HanPhP2004, Meng+2009,  Ruiter+2009}. \citet{GilfanovBogdan2010} and \citet{diStefano2010} also argue that observed populations of SD SN Ia precursors cannot account for all SN Ia, though this may be due to incorrect assumptions about the appearance of SD systems \citep{diStefano2010b}.

The lack of hydrogen is a more acute problem, which could preclude almost any SD progenitors. Observations should be able to detect very small amounts of hydrogen ($\lessapprox 0.01M_{\odot}$) but none has been seen \citep{Leonard07}. Almost the entire envelope of a red-giant donor is believed to be stripped by the shock wave from the explosion \citep[][]{Marietta+2000}; 
such stripping should lead to significant contamination of the supernova spectrum by hydrogen. In the case of a main-sequence or subgiant donor, a smaller fraction of the envelope has been predicted to be stripped. \citet{Pakmor08} showed that the simulated amount of hydrogen stripping from main-sequence donors could still be within the observational limits, though only barely; this is uncomfortable for the SD model \citep[see also][]{MengYang2010}. Here we present a scenario in which the remaining envelope mass at explosion is naturally less than the observational $0.01M_{\odot}$ limit given by \cite{Leonard07} and also very little of it can be stripped. As our model also significantly reduces the cross-section for interaction between the supernova shock and the donor star, it would unfortunately also reduce the opportunity for such interactions to be spotted in SN Ia light curves, as proposed by \citet{Kasen2010}.

Section \ref{sec:model} explains how non-degenerate donors could finish mass transfer and shrink well inside their Roche lobes by the time of explosion, section \ref{sec:pops} discusses whether it is plausible for this chain of events to occur in a significant fraction of SD SN Ia progenitors, and section \ref{sec:06X} considers to what extent SN 2006X constrains our model.

\section{Exploding after envelope contraction}
\label{sec:model}

\begin{deluxetable}{lc}
\tablecaption{\label{tab:mignit} Changes in limiting WD mass with WD rotation} 
\tablehead{Situation & Mass ($M_{\odot}$)}
\startdata
(a) Ignition of non-rotating CO WD & 1.378 \\ 
(b) Ignition of accreting WD with strong AM transport & $\approx$ 1.5 \\
(c) Max. for accreting WDs in Yoon \& Langer model & $\approx$ 2   \\
(d) Extreme stability limit of differentially rotating WD & $\gtrsim$4 
\enddata
\tablecomments{Masses are taken from: (a) \citet{Nomoto+1984}; (b) Uenishi, Nomoto \& Hachisu (2003), Saio \& Nomoto (2004); (c) Yoon \& Langer (2005); (d)  Ostriker \& Bodenheimer (1968)}
\end{deluxetable}

Our mechanism for avoiding hydrogen contamination is built on two apparently robust effects. Firstly, accretion of angular momentum can increase the mass at which a CO WD is expected to explode and allow the mass transfer phase to continue for longer than is normally assumed. Secondly, if the donor star has a giant structure, then a natural end to the mass transfer phase occurs when the envelope mass becomes so small that the envelope rapidly shrinks.  

\subsection{Postponing ignition during accretion of angular momentum}
\label{sec:AMacc}

Rotating WDs can avoid exploding or collapsing even if their masses significantly exceed the Chandrasekhar limit ($M_{\rm Ch}$).  In the extreme case of differential rotation, a white dwarf is stable to $\rm 4~{\rm M_{\odot}}$ \citep{OstrikerBodenheimer68}. More recently, \citet{YoonLanger2004, YoonLanger2005} found that SD SN Ia progenitors can reach $\rm \approx 2~{\rm M_{\odot}}$ before exploding.  This may explain SN 2003fg, which appears to have had a super-Chandrasekhar WD progenitor \citep{Howell+06}.\footnote{See also \citet{ChenLi2009}.}  Even if accretion takes $M_{\rm{WD}}$ well above $M_{\rm{Ch}}$, ignition may not occur until angular momentum is lost from or redistributed within the WD.   

In contrast to Yoon \& Langer, \citet{SaioNomoto2004} and \citet{Piro2008} argue that angular momentum transport is very efficient, and hence that the accreting WD is unlikely to be far from solid-body rotation. However, solid body rotation can also delay ignition: for example \citet{Uenishi+2003} and \citet{SaioNomoto2004} find that the ignition mass of the WD is raised by $\approx 0.1 M_{\odot}$ even in the case of efficient angular momentum transport.\footnote{In \emph{pure} solid-body rotation, the WD stability limit is 1.47$M_{\odot}$.} Table \ref{tab:mignit} summarises the WD mass limits for different models. We consider the effect of both the solid-body and differentially-rotating cases in section \ref{sec:pops}.    

There is an important caveat to invoking differential rotation and using the higher limit of $\approx 2~M_{\odot}$. Sustaining differential rotation probably takes an accretion rate of $\gtrsim 10^{-7} M_{\odot}$ \citep{YoonLanger2004, YoonLanger2005}.\footnote{Of course, if differential rotation has been established, it continues until angular-momentum is redistributed.} High enough mass transfer rates are far from guaranteed in all cases, e.g. for RG donors with low-mass cores. However, near-solid-body rotation seems sufficient for the scenario proposed here to be common \& important (see section 3).

\begin{centering}
\begin{figure}
\epsfig{file=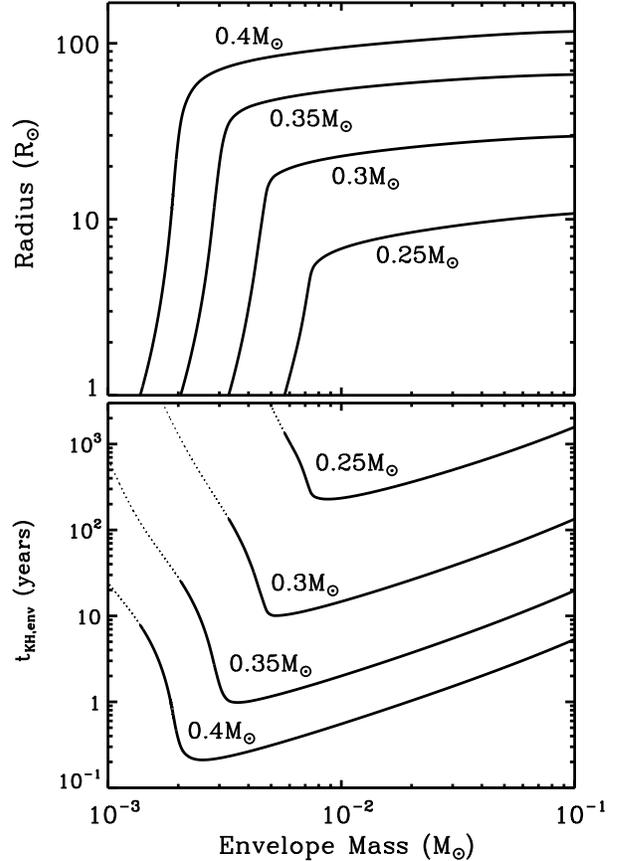, width=8cm}
\caption{\label{fig:tkhcollapse}
The upper panel shows stellar radii as a function of envelope mass; the lower panel shows the thermal timescales of those envelopes. The labels next to the the curves show the fixed core masses used for each calculation and the curves in the lower panel become dotted when the stellar radius drops below 1 $R_{\odot}$. The knee in the radius-mass plot marks where the envelope starts to shrink; in all cases, this happens with $M_{\rm env}$ below the $0.01M_{\odot}$ limit for hydrogen detection given by Leonard (2007).  }
\end{figure}
\end{centering}

\subsection{Rapid contraction of giant envelopes} 
\label{sec:collapse}

If continued accretion can prevent the explosion, then the SN Ia will be postponed until after mass transfer ceases (and then until after sufficient angular momentum of the WD has been lost or redistributed).  When the mass of a RG envelope ($M_{\rm env}$) becomes too low for the star to maintain a red giant structure then the envelope rapidly contracts on its thermal, or Kelvin-Helmholtz, timescale $\tau_{\rm KH,env}$, given by:
\begin{equation}
\label{eq:tkh}
\tau_{\rm KH,env} \approx \frac{G M_{\rm star } M_{\rm env}}{2 R L}
\end{equation}
where $G$ is the gravitational constant, $M_{\rm star }$ and $M_{\rm env}$ are the mass of the star and the envelope and $R$ and $L$ are the radius and luminosity of the star. 
We can scale that timescale to approximate stellar properties on the giant branch as:
\begin{equation}
\label{eq:tkh_scaled3}
\frac{\tau_{\rm KH,env}}{\rm 48~yr} \approx \left( \frac{M_{\rm star}}{0.3  M_{\odot}} \right) \left( \frac{M_{\rm env}}{10^{-2} M_{\odot}} \right) \left(\frac{R}{10  R_{\odot}} \right)^{-1}  \left(\frac{L}{10^{2} L_{\odot}}\right)^{-1}.
\end{equation}
Since the luminosity is a steep function of core mass, this timescale drops rapidly for donors higher up the giant branch. Figure \ref{fig:tkhcollapse} shows $\tau_{\rm KH,env}$ when the envelope shrinks, calculated for a set of $\sim$equilibrium stellar structures with different fixed core masses. We adopt a metallicity of 0.02 and use Eggleton's stellar evolution code \citep{PPE1971, Pols+95}. In reality the envelope is depleted by shell burning from below as well as by mass loss from above; also the envelope will not be in strict equilibrium during mass transfer. For each core mass, then above a certain envelope mass the radius is roughly constant; below that envelope mass the envelope shrinks rapidly.\footnote{Mass transfer is driven by the growth of the core due to nuclear burning.} 
If the donor star in RS Oph is Roche-lobe filling, it has a core mass of $\approx 0.4 M_{\odot}$ \citep[see, e.g.~][]{Justham+09}; based on these calculations, if the donor's envelope mass were to fall below $\sim 2 \times 10^{-3} M_{\odot}$, the envelope would contract in $\tau_{\rm KH,env} < 1$ yr.

Once contraction of the envelope is complete then the radius of the star is essentially insignificant; taking a separation of $\approx 100R_{\odot}$ and a core size of $\approx 0.1 R_{\odot}$ gives a fractional cross-section of $\sim 10^{-6}$ for interaction between the supernova shock and the companion star. In addition, after contraction any remaining envelope will be tightly bound to the compact core.

\subsection{The delay before explosion}
\label{sec:simmering}

After the end of accretion, the explosion will happen only after the WD redistributes or loses sufficient angular momentum (for differential or solid-body rotation, respectively). The timescale for this angular momentum loss or redistribution is uncertain. However, an upper limit to the angular momentum loss and redistribution timescales of $~10^{6}$ years has been claimed from mapping the central density at ignition to the expected nucleosynthesis \citep{YoonLanger2005}.  But even if this delay is negligible then the finite time that carbon burning takes to runaway should be long enough to allow the envelope to contract.  

After the energy generation rate from carbon burning exceeds the rate at which neutrinos can cool the core then convection currents are believed to postpone the explosion until the heating timescale becomes too short compared to the convective turnover timescale \citep[e.g.\ ][]{Arnett1969, Lesaffre+2006}. The likely $\sim 10^{3}$ year delay due to this `simmering' phase would allow time for exhausted RG envelopes to shrink, even without considering the time taken for redistribution or loss of angular momentum (see Fig.\ \ref{fig:tkhcollapse}).

\subsection{Potential extension to the super-soft channel} 
\label{sec:sss}

The above model naturally applies to SD progenitors of SN Ia with RG donor stars. Our proposed mechanism is significant even if it only applies to most such RG+WD SD systems. However, our scenario may also extend to at least some progenitors from the supersoft channel; such systems with late Case A or early Case B donors will evolve into systems with giant donors if there is enough mass remaining in the envelope, and if the WD does not explode first.  If the envelope mass is already too small, then there is no problem to solve: the donor will shrink to become a WD or core-helium burning hot subdwarf in a wide binary (until, that is, the explosion of the WD breaks up the binary).\footnote{For the conditions which divide donors which will produce a WD from those which can ignite helium in their core see, e.g., \cite{Han+2002}.}

\subsection{One exceptional case}

A class of exceptions to our model might be systems containing donor stars which are able to reach helium ignition after growing their WDs to $M_{\rm Ch}$ whilst still posessing a substantial hydrogen envelope. These donor stars would contract away from contact after igniting helium, allowing explosion to occur. Even in this case the donor star would no longer be Roche-lobe filling (by an order of magnitude or more).

\section{Population effects}

\label{sec:pops}

Despite the extended-accretion model for single degenerate progenitors described above, there exists a plausible explanation for why most SN Ia apparently explode within a narrow range of masses: the mass reservoir in the donor may usually be too small to allow significantly super-Chandrasekhar masses. Below we suggest that most SN Ia might reasonably still explode with WD masses $\lesssim 1.5 M_{\odot}$. The WD mass distribution at explosion directly relates to the question of what fraction of SD SN Ia are subject to our proposed mechanism.

\subsection{Principles \& estimates}

The remaining donor mass at explosion depends on the donor mass at the start of the accretion phase ($M_{d,i}$), the initial and final accretor masses ($M_{WD,i}$ \& $M_{WD,f}$) and the mean accretion efficiency (i.e.\ the fraction of the mass lost by the donor that the WD manages to accrete and retain).  There is considerable uncertainty in the population distribution of these: any model that could know them would have largely solved the question of which systems produce SN Ia.  

However, we can apply some constraints. The canonical limit for dynamically stable mass transfer limits the mass of a giant donor to be $\lessapprox 1.2$ times the mass of the accretor.\footnote{See, e.g., section 5 of \cite{Han+2002}.} So for an initial COWD mass of $\lesssim$ 1.0 $M_{\odot}$ then for a RG donor any final core mass of 0.2 $M_{\odot}$ or more would require a mean accretion efficiency above 50\% before the final mass of the accretor could exceed $\approx$ 1.5 $M_{\odot}$ (i.e. before we must appeal to strong differential rotation).  

Much lower -- even negative -- mass accretion efficiencies for such RG+WD recurrent novae are common in the literature \citep[see, e.g.,][]{Yaron+2005}. Furthermore \citet{ChenLi2009} find that rapidly rotating WDs can experience even lower accretion efficiencies than non-rotating WDs. 

For the supersoft channel, dynamical stability canonically restricts the mass of a radiative donor to be $\lesssim 3$ times the mass of the accretor, in which case overall accretion efficiencies $\lessapprox$ 20\% would allow our model to be widely applicable in the restrictive case of solid-body rotations.\footnote{For $M_{WD,i}$=1.0 $M_{\odot}$, then $M_{d,i} \lessapprox 3.0 M_{\odot}$; remaining a RG requires a final core mass for the donor of $\lessapprox$ 0.5 $M_{\odot}$, leaving $\lessapprox$ 2.5 $M_{\odot}$ available to transfer, of which $\lessapprox$ 20\% could be added to the WD whilst satisfying $M_{WD,f} \lessapprox 1 .5 M_{\odot}$.}  This seems challenging without a significant and inefficent recurrent-nova phase. \citet{HachisuKatoNomoto2008} have suggested that the limit for dynamical stability could be significantly higher than expected, which would make it very difficult for the mechanism presented here to apply to those SD systems.

\subsection{Population synthesis}


For a population of RG donor stars, \citet{MengYang2010} calculated remaining envelope masses at explosion under the standard assumption that SN Ia occur at a CO WD mass of 1.378 $M_{\odot}$.  Extrapolating from their results, then for accretion efficiencies $\lessapprox 0.3$, almost none of the systems in one of the two populations they synthesised would increase the WD mass by more than 0.1$M_{\odot}$ before the donor's envelope contracted.  

For more general SD systems, \citet{ChenLi2009} \emph{attempted} to make super-Chandrasekhar SN Ia by considering accretion onto rotating WDs, but found `in most cases the final masses of the white dwarfs are not significantly exceeding 1.4$M_{\odot}$'. They state that for an initial WD mass of $1.0 M_{\odot}$ and for a metallicity of 0.02 then `the maximum explosion mass of white dwarfs is always less than 1.5 $M_{\odot}$'.  Although many of their donors had substantial remaining envelopes, that work assumed that super-Chandrasekhar WDs exploded when the mass transfer rate dropped below $3 \times 10^{-7} M_{\odot} {\rm yr^{-1}}$, i.e. a criterion which estimates when differential rotation can no longer be sustained; future work should also consider that near-solid-body rotation can prevent explosion for WD masses $\lessapprox 1.5 M_{\odot}$.

So it seems plausible that our mechanism could apply to most RG+WD SD systems, whilst an extension to the majority of other SD systems, though possible, appears more difficult.  There may be a balance between predicted SD rates and remaining envelope masses, since varying parameters to increase SD SN Ia rates (e.g. increasing mass accretion efficiencies) also seems likely to increase the incidence of donors with a significant remaining envelope mass at explosion.  We encourage population synthesis studies to simultaneously investigate the SD rates, distribution of SN Ia explosion masses and the fraction of hydrogen-contaminated SN Ia for a range of different assumptions.



\section{SN 2006X}
\label{sec:06X}

The Na absorption features observed by \citet{Patat+Sci07} in SN 2006X were interpreted as evidence of a SD progenitor.  In particular, the distribution was reminiscent of a set of nova shells.\footnote{An alternative SD explanation was suggested by \citet{HachisuKatoNomoto2008}.}  Similar features have been seen in a small number of other SN Ia \citep{Blondin+2009, Simon+2009}.

\subsection{Constraints}

The interpretation of a system with a RG donor star and recent nova outbursts constrains our model, as it provides an estimate of the time between the shrinking of the envelope and the explosion in this case. \citet{Patat+Sci07} estimate that a time of $\sim 25$ years might have passed since the last nova outburst.  This timescale for envelope contraction can easily be beaten (see Fig.\ \ref{fig:tkhcollapse}), so at least there is no difficulty with proposing that the envelope has contracted since the last outburst.  However, expected simmering times are rather longer than this; it would be helpful for our model if the $\sim$25 year timescale given by Patat et al.\ is an under-estimate.\footnote{This timescale also constrains angular momentum redistribution or loss.} Those few SN Ia that show these features may simply be fine-tuned or extreme cases. Perhaps, e.g., differential rotation truncates the simmering phase: \citet{Lesaffre+2006} suggested that differential rotation could suppress the convective instability of the core and hence potentially shorten the time before the thermonuclear runaway is triggered.

\subsection{A new model}

Alternatively, the interpretation as nova shells could be incorrect. A speculative but potentially elegant explanation occurs if, for some SN Ia -- perhaps those where the donors manage to ascend highest up the giant branch -- the remaining envelope is [partially] spontaneously ejected. For core masses $\gtrapprox 0.4 M_{\odot}$, the dynamical and thermal times of the envelopes become comparable as the envelope mass decreases (for the 0.4$M_{\odot}$ core in Fig.\ \ref{fig:tkhcollapse}, $t_{\rm KH,env} \approx 0.2$ yr at contraction, with $t_{\rm dyn}\approx 0.04$ yr,\footnote{For a total mass of 0.4$M_{\odot}$, $t_{\rm dyn} \approx 0.08 (R/100 R_{\odot})^{3/2}$ yr.} whilst for a 0.45 $M_{\odot}$ core, we expect $t_{\rm KH,env} < t_{\rm dyn}$\footnote{This inequality is true when our code fails \emph{before} the envelope shrinks. Note, however, that the dominant causal timescale may be the sound-crossing time.}). It has been argued that $t_{\rm KH,env} \lesssim 10 t_{\rm dyn}$ leads to envelope ejection on the AGB, as it should lead to pulsational instability \citep{Soker2008}. In addition, RG envelopes approaching the tip of the giant branch can have binding energies which are formally positive \citep[see, e.g., ][]{Han+1994}. That combination of conditions makes an instability to pulsational envelope ejection seem very plausible.  The pulsational instability itself or shaping by the binary orbit might produce the discrete features observed by Patat et al.\footnote{The escape velocity of an appropriate star matches velocities inferred from the Patat et al.\ data; For radius $R=60 R_{\odot}$ and mass $M=0.4 M_{\odot}$, $\sqrt{2GM/R }=50 km~s^{-1}$.}

\section[]{Summary and Conclusions}

We have presented a scenario in which SD SN Ia could avoid having Roche-lobe-filling donors at the point of explosion. Rapid rotation\footnote{Perhaps, but not necessarily, differential rotation.} of the accreting WD stabilises it against ignition until after the H-rich envelope of the donor is almost exhausted and contracts. The companion at explosion is then a WD or hot subdwarf in a wide orbit.

The key uncertainty is the fraction of SD SN Ia to which our proposed scenario applies. We have argued that it could plausibly work for the majority of systems in the RG+WD channel (for which the non-detection of hydrogen seems most problematic), and perhaps for many other SD SN Ia (depending on, e.g., accretion efficiencies).

This result suggests that searches for hydrogen in SN Ia \citep{Leonard07} and attempts to see the collision between the SN shock and the donor stars in SN Ia \citep{Kasen2010} may not distinguish between SD and DD progenitors. They may also be unable to disfavour giant donors, as concluded by \cite{Hayden+2010}.
Hence it is especially important that we make the most of the information from those SN Ia which show evidence for circumstellar material.  We have speculated on a novel origin for those features.

We have also argued that, particularly if the absorption features seen in SN 2006X are due to nova shells, they could place constraints on the speed of angular momentum redistribution within, or loss from, the accreting WD and the duration of the simmering phase before explosion. Hence such observations may not just help determine the progenitors of those SN Ia, but also given an insight into the physics operating inside the doomed WD. 




\section*{Acknowledgements}
 
Thank you to the referee and to Philipp Podiadlowski \& Natasha Ivanova for
helpful discussions and comments, and to the 
China National Postdoc Fund 
(Grant No.\ 20090450005) and the National Science Foundation of 
China (Grant Nos.\ 10950110322 and 10903001) for support.
Apologies to those authors whose SN Ia work I didn't have space to mention. 

\bibliographystyle{apj}

\end{document}